\def\be{\begin{equation}} \def\ee{\end{equation}} 
\def\bea{\begin{eqnarray}} \def\eea{\end{eqnarray}} 

\documentclass[preprint,12pt]{elsarticle}

\usepackage{amssymb}
\usepackage{amsmath}

\journal{Nuclear Physics A}

\begin{document}

\begin{frontmatter}



	\title{ 
		Radiative $\alpha$ capture on $^{12}$C 
	in cluster effective field theory: short review}


\author{Shung-Ichi Ando} 

\affiliation{organization={Department of Display and Semiconductor Engineering,
and Research Center for Nano-Bio Science, 
	Sunmoon University}, 
            city={Asan-si},
	    state={Chungcheongnam-do},
            postcode={41439}, 
            country={Republic of Korea}}

\begin{abstract}
	Study of radiative $\alpha$ capture
	on $^{12}$C, $^{12}$C($\alpha$,$\gamma$)$^{16}$O, 
	in cluster effective field theory (EFT) is reviewed. 
	A low energy EFT for 
	$^{12}$C($\alpha$,$\gamma$)$^{16}$O
	at the Gamow-peak energy, $E_G=0.3$~MeV, 
	is constructed, and the theory
	is first applied to the study of elastic $\alpha$-$^{12}$C scattering
	at low energies. 
	The effective range parameters are fitted to 
	the precise phase shift data of the elastic scattering and 
	the astrophysical $S_{E1}$ factor of 
	the $E1$ transition of $^{12}$C($\alpha$,$\gamma$)$^{16}$O 
	at $E_G$ is estimated. 
	For the study of the $E2$ transition of 
	$^{12}$C($\alpha$,$\gamma$)$^{16}$O, we discuss a difficulty to 
	determine the asymptotic normalization coefficient (ANC) 
	of the subthreshold $2_1^+$ state of $^{16}$O from 
	the elastic scattering data,
	and demonstrate the difficulty with the estimate of 
	the astrophysical $S_{E2}$
	factor of $^{12}$C($\alpha$,$\gamma$)$^{16}$O at $E_G$. 
	We discuss 
	the uncertainty in the estimate of the 
	$S$ factors at $E_G$ in the present approach. 
\end{abstract}



\begin{keyword}
	$^{12}$C($\alpha$,$\gamma$)$^{16}$O \sep
	astrophysical $S$ factor \sep 
	elastic $\alpha$-$^{12}$C scattering \sep
	asymptotic normalization coefficients \sep 
	cluster effective field theory



\end{keyword}

\end{frontmatter}



\section{Introduction}
\label{sec1}

Radiative $\alpha$ capture on $^{12}$C, $^{12}$C($\alpha$,$\gamma$)$^{16}$O,
is one of the fundamental reactions in nuclear astrophysics,
which determines the C/O ratio in the core of a helium-burning 
star~\cite{f-rmp84}.
The reaction rate, equivalently the astrophysical $S$ factor of 
$^{12}$C($\alpha$,$\gamma$)$^{16}$O at the Gamow-peak energy, $E_G=0.3$~MeV,
has not been measured in an experimental facility because of the Coulomb
barrier. 
One needs to employ a theoretical model, 
fit the model parameters to the experimental
data measured at a few MeV energy, and extrapolate the cross section to $E_G$.
It is known that $E1$ and $E2$ transitions are dominant due to the subthreshold
$1_1^-$ and $2_1^+$ states of $^{16}$O. 
During the last five decades, many experimental and theoretical works 
have been carried out. See, e.g., Refs.~\cite{bb-npa06,detal-rmp17} for review. 

For the last decade, we have been studying the related reactions 
to $^{12}$C($\alpha$,$\gamma$)$^{16}$O employing the methodology of 
field theory~\cite{w-pa79,hkk-rmp20,dgh-14}. 
In constructing a theory, one first needs 
to choose a typical scale of a reaction to study. 
We choose the Gamow-peak energy, $E_G=0.3$~MeV, as a typical energy scale, 
and thus a typical 
momentum scale would be $Q=\sqrt{2\mu E_G} = 40$~MeV/c where $\mu$ is
the reduce mass of $\alpha$ and $^{12}$C; the typical length scale becomes
$\hbar/Q \simeq 5$~fm. The reaction would be insensitive to the 
nucleon degrees of freedom inside of the nuclei, and we treat the $\alpha$ 
and $^{12}$C as point-like scalar fields.
One also chooses a large scale to separate relevant degrees of freedom
at low energies from irrelevant degrees of freedom at high energies. 
We choose the energy difference between the threshold energies of 
the $p$-$^{15}$N and $\alpha$-$^{12}$C channels of $^{16}$O, 
$\Delta E = 12.13-7.16=4.79$~MeV, as the high energy (separation) scale;
the high momentum scale is $\Lambda_H = \sqrt{2\mu \Delta E} = 160$~MeV/c. 
The theory provides us with a perturbative expansion scheme and the expansion
parameter would be $Q/\Lambda_H = 1/4$. 
The $p$-$^{15}$N system is now regarded as the irrelevant degrees of freedom
and integrated out of the effective Lagrangian,
whose effects are embedded in the coefficients of terms of the Lagrangian. 
Those coefficients can, in principle, be determined from the mother theory,
while they, in practice, are fixed by using experimental data or empirical 
values of them. 
Because of the perturbative expansion scheme of EFT, truncating the terms
up to a given order, one can have an expression of reaction amplitudes
in terms of a few parameters for each of the reaction channels. 

The $R$-matrix analysis and two-body potential models
                are standard methods to analyze the $\alpha$-$^{12}$C system
		(see Table IV in Ref.~\cite{detal-rmp17}).
		Because the reactions of the 16-nucleon system must be reduced
                in a system with a small number of degrees of freedom,
                model (or scheme) dependence of those methods is unavoidable.
The reduction to lower dimensions of the degrees of freedom of the reactions
may result in
the different parameterizations of these theoretical approaches.
                Differences between the $R$-matrix analysis and EFT is that the
                $R$-matrix analysis introduces a radius parameter 
		to separate the inner and outer
                parts of the wavefunctions, along with 
		a summation of resonant poles,
                the $R$ matrix, where the resonant energies and reduced widths
                in the $R$ matrix are fitted to the experimental data.
                EFT introduces a separation scale in the momentum space and
                employs the effective range expansions.
                A difference between the two-body potential models and EFT is
that the potential models employ a wave picture in coordinate
                space where the potentials are parameterized
                as a Woods-Saxon form and the parameters of the potential
                and the effective charges in the transition operators are
                fitted to the experimental data,
                while EFT adopts a particle picture in momentum space where
                the reaction amplitudes are calculated from
                the effective Lagrangian and the coupling constants of the
                vertex functions are fitted to the experimental data.
                EFT is, therefore, a new and alternative approach 
		for the study of
                the $\alpha$-$^{12}$C system and
                can provide a method
                to study the model (or scheme) dependence of
                the calculations of the $\alpha$-$^{12}$C system.

In this contribution, we briefly discuss the studies of 
$^{12}$C($\alpha$,$\gamma$)$^{16}$O in the cluster EFT.
We first discuss the calculation of the $E1$ transition amplitudes
and how the parameters are fitted to the experimental data.
We then extrapolate the $S_{E1}$ factor of 
$^{12}$C($\alpha$,$\gamma$)$^{16}$O to $E_G$. 
For the $E2$ transition amplitudes of $^{12}$C($\alpha$,$\gamma$)$^{16}$O,
we discuss the difficulty of fixing the effective 
range parameters (equivalently the ANC) 
of the subthreshold $2_1^+$ state of $^{16}$O from 
the phase shift of elastic $\alpha$-$^{12}$C scattering for $l=2$ and
its consequence of the estimate of $S_{E2}$ factor of 
$^{12}$C($\alpha$,$\gamma$)$^{16}$O at $E_G$. We also discuss 
the uncertainty of persistent 
to deduce the $S$ factors of $^{12}$C($\alpha$,$\gamma$)$^{16}$O
at $E_G$ in the theory.  

\section{Calculation and numerical results}
\label{sec2}

\begin{figure}[t]
\centering
	\includegraphics[width=0.5\textwidth]{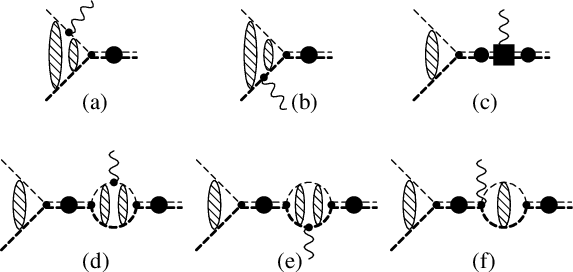}
	\caption{Feynman diagrams of the $E1$ and $E2$ transitions of 
	$^{12}$C($\alpha$,$\gamma$)$^{16}$O~\cite{sa-24}. 
	A thin or thick dashed line denotes the incoming $\alpha$ or $^{12}$C,
	and a wavy line does the outgoing photon. A thin-thick double-dashed 
	line with a filled circle in the final state or in the intermediate
	state denotes the outgoing ground state of $^{16}$O or 
	the dressed $^{16}$O propagator for $l=1$ or $l=2$.
	A shaded oval in the initial state or in the intermediate state
	denotes the Coulomb wavefunction or Coulomb greens function. 
	A filled box in diagram (c) denotes the counter terms by which 
	infinities from the loop diagrams are renormalized. }
	\label{fig1}
\end{figure}

In Fig.~\ref{fig1},
Feynman diagrams of the $E1$ and $E2$ transitions of 
$^{12}$C($\alpha$,$\gamma$)$^{16}$O are displayed, where the initial
$\alpha$-$^{12}$C states are $p$-wave and $d$-wave states, 
respectively~\cite{sa-prc19,sa-24}.
Up to the sub-leading order, the transition amplitudes can be described
by five parameters for each of the channels. 
One is the coupling constant $y^{(0)}$ of the constant vertex 
for the transition of 
$s$-wave $\alpha$-$^{12}$C state to 
the ground $0_1^+$ state of $^{16}$O, 
which is related to the ANC of the
ground state of $^{16}$O; it appears as an overall coefficient in all
of the amplitudes. 
The second one is the coefficient of the contact $O^*\gamma O$ vertex,
$h_R^{(1)}$ or $h_R^{(2)}$, in the diagram (c), which renormalizes the 
infinities from the loop diagrams in diagrams (d), (e), (f). 
The remaining three parameters are the effective range parameters 
appearing in the dressed $^{16}$O propagators in diagrams (c), (d),
(e), (f). Those parameters are fitted to the 
experimental data. 

We first fix the effective range parameters in the dressed $^{16}$O 
propagators for $l=0,1,2,3$. We constructed the $S$ matrices of the 
elastic $\alpha$-$^{12}$C scattering for $l=0,1,2,3,4,5,6$ at low energies,
where four parameters, in general, are introduced 
for each of the bound and resonant states of $^{16}$O~\cite{sa-prc23} 
and are fitted to the precise phase shift 
data~\cite{tetal-prc09}. We found that the fitted parameters reproduce
the precise phase shift data very well.\footnote{
	The sub-threshold $1_1^-$ and resonant $1_2^-$ 
states of $^{16}$O can be described by a single $^{16}$O propagator
with the effective range parameters \{$r_1$, $P_1$, $Q_1$\}~\cite{sa-prc23}, 
and, thus, there is no interference between the two states. } 
\begin{figure}
\centering
	\includegraphics[width=0.5\textwidth]{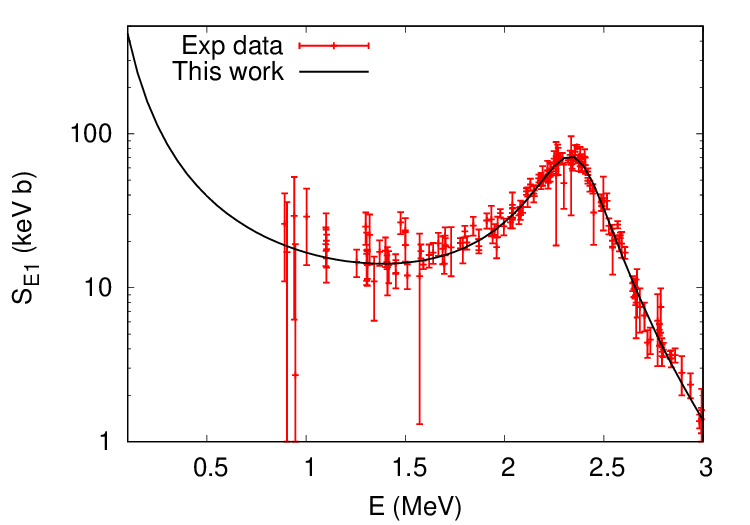}
	\caption{A fitted line of the $S_{E1}$ factor of 
	$^{12}$C($\alpha$,$\gamma$)$^{16}$O plotted as a function of
	the energy $E$ of the initial $\alpha$-$^{12}$C state in the 
	center-of-mass frame. 
	The experimental data of the $S_{E1}$ factor are displayed in the 
	figure as well. }
	\label{fig2}
\end{figure}
By using the fitted values of the effective range parameters for $l=1$,
we fit the two additional parameters of the $E1$ transition amplitudes
of $^{12}$C($\alpha$,$\gamma$)$^{16}$O to the experimental data of $S_{E1}$
factor of $^{12}$C($\alpha$,$\gamma$)$^{16}$O by employing the dimensional
regularization for the loop integrals, 
and we have the fitted values of the parameters as 
$y^{(0)} = 0.260$~MeV$^{-1/2}$ and $h_R^{(1)}=2.67\times 10^4$~MeV$^3$ 
with the $\chi^2$ value, $\chi^2/N = 1.73$, where $N$ is the number of 
the experimental data. In Fig.~\ref{fig2}, we plot the $S_{E1}$ factor
as a function of the energy $E$ of the initial $\alpha$-$^{12}$C state
in the center-of-mass frame. The experimental data are displayed in the
same figure as well. Thus, a value of the $S_{E1}$ factor of 
$^{12}$C($\alpha$,$\gamma$)$^{16}$O at $E_G$ is obtained as
\bea
S_{E1} = 59\pm 2 \ \ \textrm{keV\,b}\,,
\eea
where we have a small, about 3\,\% error bar (and a relatively large 
$\chi^2$ value). 
This mainly stems from the fact that the fitted effective range parameters 
for $l=1$ are accurate and control the energy dependence of the $S_{E1}$ 
factor in the whole energy region displayed in the figure. 
A large systematic uncertainty may appear from the fitted value of $y^{(0)}$.
Though we fitted it to the experimental
data of the $S_{E1}$ factor, it is related to the ANC of ground $0_1^+$ 
state of $^{16}$O; the value of the ANC of the ground state is still not
known well. (See, e.g., Table 3 in Ref.~\cite{sa-fbs24}.)
In addition, it also appears in the $E2$ transition amplitudes of 
$^{12}$C($\alpha$,$\gamma$)$^{16}$O. 
We will see below that the fitted value of $y^{(0)}$ to 
the data of the $S_{E2}$ factor is significantly different from that fitted 
to the data of the $S_{E1}$ factor. 

\begin{figure}
\centering
	\includegraphics[width=0.49\textwidth]{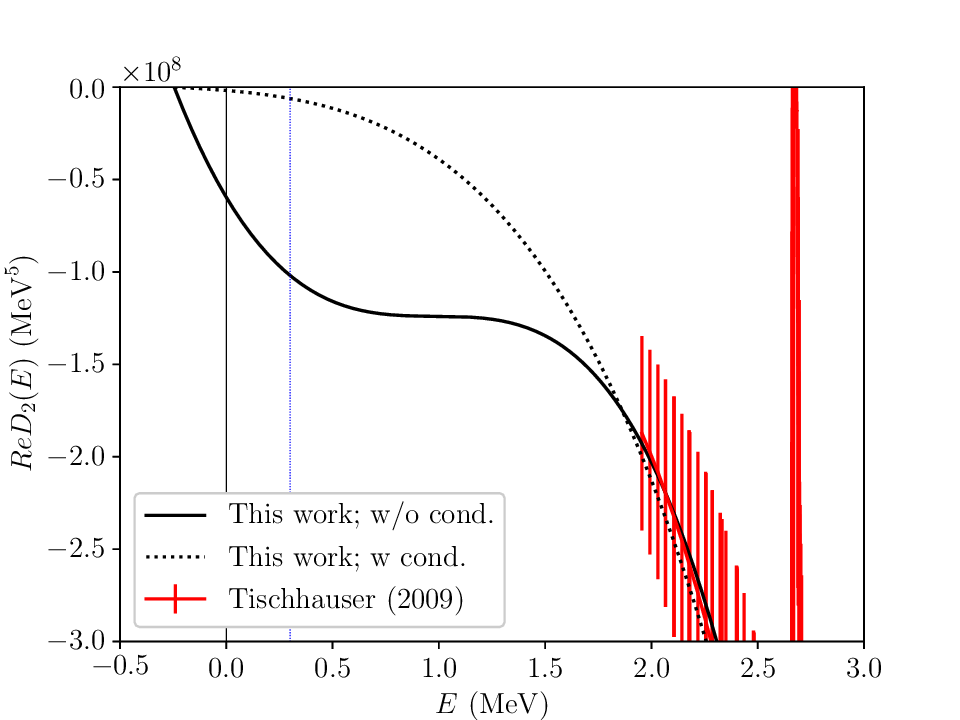}
	\includegraphics[width=0.49\textwidth]{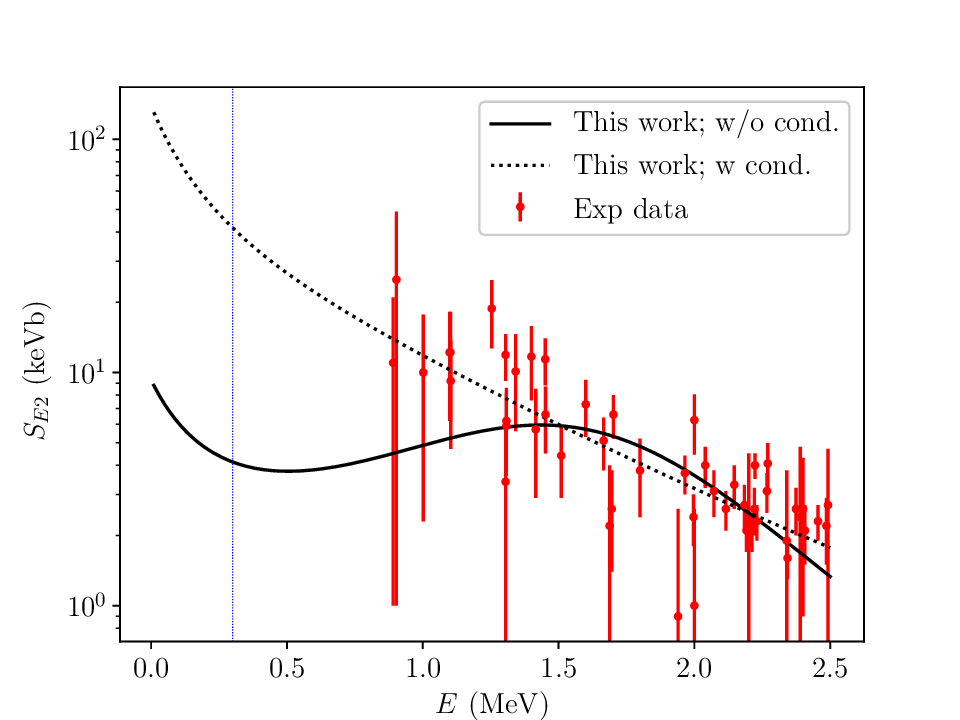}
	\caption{(left panel) A solid or dotted line of the real part of 
	the inverse of the dressed $^{16}$O propagator 
	for $l=2$, $ReD_2(E)$, 
	not applying the conditions (discussed in the text) 
	or applying the conditions plotted 
	as functions of the energy $E$ of the initial $\alpha$-$^{12}$C state 
	in the center-of-mass frame. 
	The experimental phase shift data are displayed in the 
	figure as well. A vertical blue dotted line is drawn at $E=E_G$. 
	(right panel) A solid or dotted line of the $S_{E2}$ factor of 
	$^{12}$C($\alpha$,$\gamma$)$^{16}$O plotted as functions of $E$. 
	The experimental data of the $S_{E2}$ factor are displayed in the 
	figure as well. 
	See the text for details.  
	(The figures are originally presented in Ref.~\cite{sa-24}.)
	}
	\label{fig3}
\end{figure}

A known problem to deduce the ANC of the subthreshold $2_1^+$ state of $^{16}$O
is that the values of the ANC deduced from the phase shift data of the 
elastic scattering is about a factor of five smaller than those deduced
from the $\alpha$ transfer reactions. 
This feature may be seen in the 
plot of the real part of the inverse 
of the dressed $^{16}$O propagator for $l=2$, $ReD_2(E)$.
(We will discuss the parameter fit of $ReD_2(E)$ in the following.)
In the left panel of Fig.~\ref{fig3}, we plot two lines of $ReD_2(E)$ as 
functions of the energy $E$ of the $\alpha$-$^{12}$C state 
at the small energy region where the $S_{E2}$ factor
is extrapolated to $E_G$. Both the lines start at the same point 
(at the upper left side of the figure) where 
$D_2(E=-B_2)=0$: $B_2$ is the binding energy of the $2_1^+$ state of $^{16}$O.
The gradients of the lines at the point are related to the ANC of the 
$2_1^+$ state; a large negative angle corresponds to a small value of the ANC,
$|C_b|_2 = 3.24\times 10^4$~fm$^{-1/2}$,
and a small negative angle does to a large value of the ANC,
$|C_b|_2=22.8\times 10^4$~fm$^{-1/2}$. 
Here we expand $ReD_2(E)$ around $E=-B_2$ as
\bea
ReD_2(E) \simeq \sum_{n=1}^5C_n(E+B_2)^n\,,
\eea
and introduce the conditions, $C_n <0$ with $n=1,2,3$ when fitting the 
effective range parameters to the phase shift data to obtain a
smoothly decreasing line of $ReD_2(E)$ (the dotted lines in the two figures
in Fig.~\ref{fig3})~\cite{sa-24}. 
The two lines go through
the quite different points at $E=E_G$ while both the lines reproduce 
the precise phase shift data reported by Tischhauser et al.
(2009)~\cite{tetal-prc09} 
equally well.
Thus, while the gradient of the line at the point, i.e., the ANC, is 
the important constraint to extrapolate the $S_{E2}$ factor to $E_G$, 
the precise phase shift data are not useful
to determine the ANC of the $2_1^+$ state of $^{16}$O. 

By employing 
the two lines of $ReD_2(E)$ displayed in the left
panel of Fig.~\ref{fig3} in the $E2$ transition amplitudes of 
$^{12}$C($\alpha$,$\gamma$)$^{16}$O, we fit the additional two parameters,
$y^{(0)}$ and $h_R^{(2)}$, to the experimental data of the $S_{E2}$ factor
of $^{12}$C($\alpha$,$\gamma$)$^{16}$O below the energy of the sharp resonant
$2_2^+$ state of $^{16}$O. 
We note that we refit the dotted line of $ReD_2(E)$ by using a value
of the ANC, $|C_b|_2 = 10\times 10^4$~fm$^{-1/2}$, which is about the center 
between the small and large values of the ANC and close to those deduced
from the $\alpha$ transfer reactions.
Thus, we have 
$y^{(0)}= 1.99\pm 0.01 \times 10^{-3}$~fm$^{-1/2}$ and 
$h_R^{(2)}= 50.6\pm 0.4 \times 10^{11}$~MeV$^4$ 
with $\chi^2/N=1.55$ for the small value of the ANC, 
$|C_b|_2=3.24\times 10^4$~fm$^{-1/2}$, and 
$y^{(0)}= 5.8 \pm 0.1 \times 10^{-2}$~fm$^{-1/2}$ and 
$h_R^{(2)}= 45.53^{+0.04}_{-0.03} \times 10^{11}$~MeV$^4$ 
with $\chi^2/N=1.18$ for a large value of the ANC, 
$|C_b|_2 = 10\times 10^4$~fm$^{-1/2}$. 
In the right panel of Fig.~\ref{fig3}, we plot the two lines of the $S_{E2}$ 
factor as functions of $E$.
The experimental data of the $S_{E2}$ factor are displayed in the figure 
as well. 
One may find that the energy dependencies of the two lines mainly stem 
from the lines of $ReD_2(E)$
plotted in the left panel of Fig.~\ref{fig3}. Thus, it would be critical 
to have the right values of the effective range parameters for $l=2$ 
at the small energy region, less than 2.0~MeV: we have the $S_{E2}$ factor
at $E_G$ as
\bea
S_{E2} &=& 
42^{+14}_{-13}\ \ \textrm{keV\,b}\,, 
\eea
for the ANC, $|C_b|_2 = 10\times 10^4$~fm$^{-1/2}$. 
(We have $S_{E2}= 4.1\pm 0.2~\textrm{keV b}$ at $E_G$ for the small ANC.)
We have a large, about 33\% error of the $S_{E2}$ factor with a small 
$\chi^2$ value, $\chi^2/N = 1.18$, for a realistic value of the ANC,
$|C_b|_2 = 10\times 10^4$~fm$^{-1/2}$. 
The large error bar may be caused by the 
scattered data with large errors of the $S_{E2}$ factor.
As discussed above, the estimate of the $S_{E2}$ factor at $E_G$ also 
contains a systematic uncertainty because the coupling 
constant $y^{(0)}$ is fitted to the data of the $S_{E2}$ factor and 
the fitted value of $y^{(0)}$ is quite different from that in the case of 
the $S_{E1}$ factor. 

\section{Summary and discussion}
\label{sec3}

In the present contribution paper, we briefly reviewed the studies of 
the $E1$ and $E2$ transitions of $^{12}$C($\alpha$,$\gamma$)$^{16}$O and the 
estimates of the astrophysical $S_{E1}$ and $S_{E2}$ factors of 
$^{12}$C($\alpha$,$\gamma$)$^{16}$O at $E_G$ in the cluster EFT.
We found the small statistical error of the $S_{E1}$ factor 
and the large statistical one of the $S_{E2}$ factor at $E_G$.
Those features mainly stem from the fact that the extrapolations 
essentially depend on the effective range parameters for $l=1$ and $l=2$.
The effective range parameters for $l=1$ are accurately constrained by 
the phase shift data, while those for $l=2$ are not. 
We demonstrated the large uncertainty of the $S_{E2}$ factor extrapolated 
to $E_G$ for the two cases, the small and large values of the ANC of the 
$2_1^+$ state of $^{16}$O. 

By adopting a realistic value of the ANC, 
we constrain the effective range parameters for $l=2$
and estimate the $S_{E2}$ factor at $E_G$.
We find about 33\% error of the $S_{E2}$ factor at $E_G$ due to the 
large uncertainty from the data of the $S_{E2}$ factor. 
Another source of the uncertainties is the value of $y^{(0)}$.
$y^{(0)}$ is related to the ANC of the ground state of $^{16}$O,
which is not well defined because the separation scale of $\alpha$-$^{12}$C
state in the ground state of $^{16}$O is smaller than the size of $^{16}$O.
In addition, as discussed above, the values of $y^{(0)}$ are still largely 
scattered
in the cases of the fit to the experimental data of the $S$ factors.
Therefore, the parameter $y^{(0)}$ may remain in the main uncertainty of the
estimates of the $S$ factors of $^{12}$C($\alpha$,$\gamma$)$^{16}$O at 
$E_G$ in the theory.  

\section*{Acknowledgments}

The present work was supported by the National Research Foundation of 
Korea (NRF) grant funded by the Korean government (MSIT) 
(No. 
2022R1F1A1070060 and 2023R1A2C1003177) and the Korean Evaluation 
Institute of Industrial Technology (KEIT) grant funded by the Korean government
(MOTIE) (No. 20022473). 








\end{document}